\documentclass[10pt, conference, compsocconf]{IEEEtran}

\IEEEoverridecommandlockouts
\usepackage[utf8]{inputenc}
\usepackage[normalem]{ulem}
\usepackage{cite}
\usepackage{amsmath,amssymb,amsfonts,amsthm}
\usepackage{algorithmic}
\usepackage{graphicx}
\usepackage{textcomp}
\usepackage{xcolor}
\usepackage{array}
\usepackage{booktabs}
\PassOptionsToPackage{bookmarks=false}{hyperref}
\usepackage[
  colorlinks = true,
  linkcolor = blue,
  urlcolor  = blue,
  citecolor = blue,
  anchorcolor = blue]{hyperref}
\def\BibTeX{{\rm B\kern-.05em{\sc i\kern-.025em b}\kern-.08em
    T\kern-.1667em\lower.7ex\hbox{E}\kern-.125emX}}

\newcommand{\N}{\mathbb{N}}

\newcommand{\E}{\mathbb{E}}
\newcommand{\R}{\mathbb{R}}

\renewcommand{\phi}{\varphi}

\newcommand{\fanin}{f_\text{in}}
\newcommand{\fanout}{f_\text{out}}
\newcommand{\fanoutleader}{f_\text{out}^\text{leader}}
\newcommand{\TTLdirect}{\mathit{TTL}_\text{direct}}
\newcommand{\TTLdigest}{\mathit{TTL}}
\newcommand{\TTL}{\mathit{TTL}}
\newcommand{\tpush}{t_\text{push}}
\newcommand{\tpull}{t_\text{pull}}
\newcommand{\tantientropy}{t_\text{recovery}}

\newboolean{showcomments}
\setboolean{showcomments}{false}
\ifthenelse{\boolean{showcomments}}
{ \newcommand{\mynote}[3]{
   \fbox{\bfseries\sffamily\scriptsize#1}
   {\small$\blacktriangleright$\textsf{\emph{\color{#3}{#2}}}$\blacktriangleleft$}}}
{ \newcommand{\mynote}[3]{}}

\newcommand{\vspacebeforecaption}{\vspace{-5mm}}
\newcommand{\vspaceaftercaption}{\vspace{-2mm}}

\begin{document}

\title{Fair and Efficient Gossip in Hyperledger Fabric\\
\small{This paper was accepted for publication in the IEEE ICDCS 2020 conference (\url{https://icdcs2020.sg}); The copyright is with the IEEE.}
}

\author{%
\IEEEauthorblockN{Nicolae Berendea$^{\star}$,
Hugues Mercier$^{\diamond}$,
Emanuel Onica$^{\star}$ and
Etienne Rivière$^{\bullet}$\smallskip}
\IEEEauthorblockA{%
$^{\star}$~Faculty of Computer Science, Alexandru Ioan Cuza University, Iași, Romania
--
\href{mailto:eonica@info.uaic.ro}{\{nicolae.berendea, eonica\}@info.uaic.ro} \\
$^{\diamond}$~Institute of Computer Science, Université de Neuchâtel, Switzerland
--
\href{mailto:{hugues.mercier@unine.ch}}{hugues.mercier@unine.ch} \\
$^{\bullet}$~ICTEAM, UCLouvain, Belgium
--
\href{mailto:etienne.riviere@uclouvain.be}{etienne.riviere@uclouvain.be}
}}


\maketitle




\begin{abstract}
	
	Permissioned blockchains are supported by identified but individually untrustworthy nodes, collectively maintaining a replicated ledger whose content is trusted.
	The Hyperledger Fabric permissioned blockchain system targets high-throughput transaction processing.
	Fabric uses a set of nodes tasked with the ordering of transactions using consensus.
	Additional peers endorse and validate transactions, and maintain a copy of the ledger.
	The ability to quickly disseminate new transaction blocks from ordering nodes to all peers is critical for both performance and consistency.
	Broadcast is handled by a gossip protocol, using randomized exchanges of blocks between peers. 

	We show that the current implementation of gossip in Fabric leads to heavy tail distributions of block propagation latencies, impacting performance, consistency, and fairness.
	We contribute a novel design for gossip in Fabric that simultaneously optimizes propagation time, tail latency and bandwidth consumption. 
	Using a 100-node cluster, we show that our enhanced gossip allows the dissemination of blocks to all peers more than 10 times faster than with the original implementation, while decreasing the overall network bandwidth consumption by more than 40\%.
	With a high throughput and concurrent application, this results in 17\% to 36\% fewer invalidated transactions for different block sizes.
\end{abstract}





\vspace{-1mm}
\section{Introduction}
\label{sec:introduction}

Blockchains have gained strong momentum in the last decade, following the initial release of the Bitcoin cryptocurrency~\cite{bitcoin_nakamoto}.
A blockchain is an immutable, append-only and globally replicated data structure.
The content of the blockchain evolves following the evaluation of \emph{transactions} submitted by its clients.
Transactions who meet validity and determinism criteria are eventually executed in some unique order.
Transactions are grouped into blocks, which are linked as a chain:
each new block's header contains a cryptographic hash of the previous block, repeatedly until the first (genesis) block, allowing to check the consistency of the chain and that of any of its blocks.
A set of nodes support the blockchain infrastructure.
The key benefit of blockchains is that clients do not need to trust individual nodes or any other client, but can put their trust in the infrastructure as a whole:
The blockchain is a mutual source of agreement between otherwise mutually untrusting participants.
Blockchains enable a number of new applications, starting from cryptocurrencies~\cite{bitcoin_nakamoto,eyal2017blockchain} and followed by supply chain management~\cite{francisco2018supply}, healthcare management~\cite{kuo2017blockchain} and e-voting~\cite{kshetri2018blockchain}.

A blockchain infrastructure is a multi-layer system that combines several key enabling features~\cite{dinh2018untangling}. 
Transactions can be written in a domain-specific language (e.g. for financial transactions in bitcoin~\cite{bitcoin_nakamoto}) or a more general language (e.g. the Turing-complete and deterministic Solidity in Ethereum~\cite{wood2014ethereum} or general-purpose languages such as Go or JavaScript in Hyperledger Fabric~\cite{Hyperledger_EuroSys18}).
A runtime allows executing transactions against the current state of the ledger (i.e., the sequence of all validated transactions) using a specific virtual machine~\cite{hildenbrandt2018kevm} or container-based isolation.
Cryptographic primitives allow checking the validity of the blocks and of the blockchain itself.

In addition to language, runtime and cryptography building blocks, a blockchain relies on two key distributed protocols: \emph{consensus} and \emph{broadcast}.

Consensus manages the addition of new blocks to the chain.
As individual infrastructure nodes are generally untrustworthy and may attempt to corrupt the content of the chain or impose ill-formed blocks, byzantine fault-tolerant (BFT) consensus is generally necessary. In some scenarios, a crash-fault-tolerant (CFT) consensus might be sufficient.
%
In \emph{Open} or \emph{permissionless} blockchains, exemplified by Bitcoin~\cite{bitcoin_nakamoto}, 
Ethereum~\cite{wood2014ethereum},
or Zerocash~\cite{ben2014zerocash}, any node, called a \emph{miner}, may participate to the BFT consensus and attempt to append new blocks to the chain. 
There is no management of identities, and therefore the BFT consensus protocol must be resilient to sybil attacks~\cite{Douceur_Sybil_IPTPS01}.
%
\emph{Permissioned} blockchains, on the other hand, assume the existence of a trusted membership management service, and that nodes participating to the infrastructure are all certified by this authority.
This allows the use of classical byzantine fault-tolerant consensus algorithms that assume identifiable participants, called \emph{orderers}, such as PBFT~\cite{pbft_tocs_02}, Zyzzyva~\cite{zyzzyva_sosp_07}, or BFT-Smart~\cite{bftsmart_dsn_14}, or CFT protocols such as primary-backup replication~\cite{hunt2010zookeeper,kafka}.
Examples of permissioned blockchains are 
Hyperledger Fabric~\cite{Hyperledger_EuroSys18}, 
Tendermint~\cite{buchman2016tendermint}
and Corda~\cite{s3_corda}.

The broadcast primitive is used at multiple levels of a blockchain and is fundamental for both performance and reliability.
It propagates transactions from clients to miners in open blockchains, and new blocks from miners or orderers to all other nodes.
Since blockchains can be supported by hundreds of nodes, the direct sending of a block from its source to all other nodes is not a valid option.
Furthermore, blockchains are expected to work under challenging conditions such as churn, packet loss, and lack of synchronicity.
Deterministic algorithms perform poorly and do not scale well under such conditions, thus robust broadcast in blockchains relies on the use of randomized gossip protocols~\cite{demers1988epidemic,boyd2006randomized}.
Gossip spreads information through probabilistic exchanges between nodes, similar to the spread of an epidemic. 

\smallskip
\noindent
\textbf{Motivation and contributions.}
%
%
While the performance and scalability of byzantine fault-tolerant consensus in blockchains have received significant attention, the performance and scalability of broadcast, and its impact on the overall performance and reliability of a blockchain, has received little attention.
The objective of this work is to close this gap, with an emphasis on permissioned blockchains.
We analyze the performance and impact on reliability and consistency of gossip-based broadcast in Hyperledger Fabric~\cite{Hyperledger_EuroSys18}, and propose, implement and test protocol modifications to reduce this impact.
Our experimental validation using a 100-node cluster shows that our enhanced gossip module allows the dissemination of blocks to all peers more than 10 times faster than with the original implementation, while decreasing the overall network bandwidth consumption by more than 40\%.
This results in 17\% to 36\% fewer invalidated transactions at validation time, depending on the block generation period.
Remaining conflicts are a result of the delay of consensus and validation, henceforth our enhanced gossip module essentially eliminates most conflicts resulting from inefficient or unfair dissemination. Our code is released as open source and available at \url{https://github.com/berendeanicolae/fabric/tree/fair-and-efficient-gossip}.

\vspace{-1mm}
\section{The anatomy of Hyperledger Fabric}
\label{section:background}

\begin{figure}[t!]
	\centering
	\includegraphics[scale=0.75]{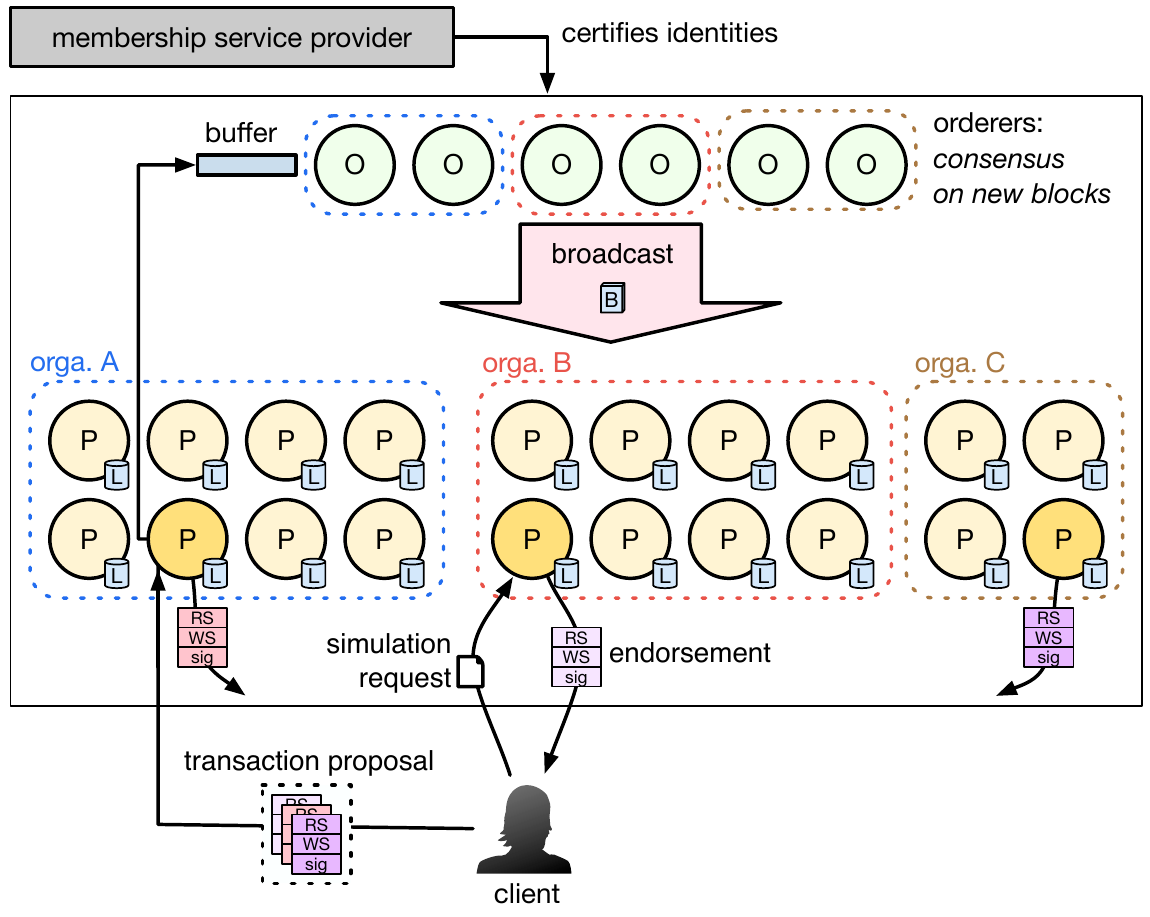}
	\vspacebeforecaption
	\caption{Fabric deployed over three organizations.\vspaceaftercaption}
	\label{fig:architecture_fabric}
	\vspace{-3mm}
      \end{figure}

Hyperledger is a collection of open-source blockchain projects initiated in December 2015 by the Linux foundation and supported by several key industrial players.
Fabric~\cite{Hyperledger_EuroSys18} is a permissioned blockchain system part of Hyperledger.
In this section we first present the architecture and components of Fabric.  
We then explain how Fabric processes client transactions, and discuss how its consistency model can result in conflicts and discarded transactions.
Note that the gossip-based broadcast of Fabric, as well as its impact on consistency, are at the core of this work and detailed in Section~\ref{section:gossip_in_fabric}.

\subsection{Fabric architecture}

The architecture of Fabric is illustrated in Figure~\ref{fig:architecture_fabric}.
Fabric supports smart contracts named \emph{chaincodes} submitted by clients and executed against the content of the ledger.
Chaincodes can be written in general-purpose languages such as node.js or Go.
As for other blockchains, a chaincode must be \emph{deterministic}:
for a given input set, the set of written values should be unique across executions.
This property enables the execution and validation of the chaincode on multiple, mutually untrusted nodes.
The Fabric execution model is not linked to a cryptocurrency.

Fabric assumes the existence of a trusted membership service provider (MSP) that uniquely certifies the identity of infrastructure nodes composed of \emph{orderers} and \emph{peers}. 
Fabric partitions the introduction of new blocks to the chain, which is handled by orderers, from the validation and execution of chaincodes, handled by peers.
This partition enables orderers to use pluggable consensus protocols such as a crash fault-tolerant service based on Apache Kafka~\cite{kafka}, or a byzantine-fault tolerant service using BFT-Smart~\cite{bftsmart_dsn_14,sousa2018byzantine}.

Fabric adopts an \emph{execute-order-validate} (EOV) model for transaction handling and execution by peers, who all maintain a complete copy of the ledger.
A subset of peers called \emph{endorsers} first performs an initial execution (called \emph{simulation}) of chaincodes.
The client collects endorsements and combines them into a transaction proposal.
Proposals are submitted to the orderers, who append them to a block, eventually added to the blockchain as a result of the consensus operation. 
Every new block is then broadcast to all peers where it is locally \emph{validated}.
The output of valid transactions is then integrated into the local copies of the ledger.
Note that this EOV model is in contrast with all open blockchains, for which all transactions are executed on all peers to form tentative blocks, and for which all peers participate to the consensus, leading to an essentially sequential execution model.

\subsection{Execution of a transaction in Fabric}

\begin{figure}[t!]
	\centering
	\includegraphics[scale=0.75]{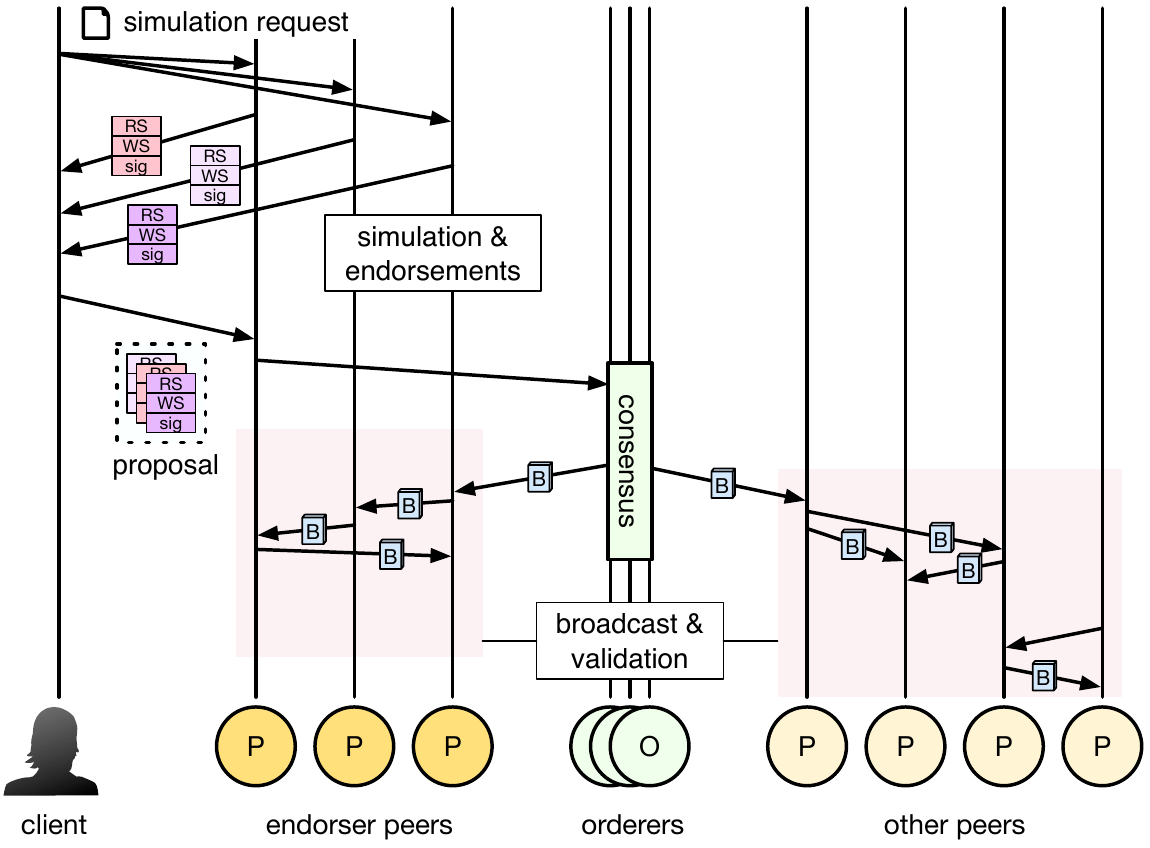}
	\vspacebeforecaption
	\caption{Timeline of a transaction preparation and execution in Fabric, with the subsequent gossip-based broadcast and validation of a new block.\vspaceaftercaption}
	\label{fig:transaction_simple}
\vspace{-3mm}
      \end{figure}

We detail the lifetime of a transaction in Fabric, from the submission of a chaincode by a client to the execution of the resulting transaction on the blockchain.
Fabric supports multiple channels (independent ledgers) and deployment over multiple organizations, as exemplified by Figure~\ref{fig:architecture_fabric}.
In the remainder of this work, we consider a deployment with a unique blockchain (channel).

Figure~\ref{fig:transaction_simple} details the timeline of transaction preparation and processing.
A client submits her chaincode to a set of peers that simulate the transaction and act as \emph{endorsers}.
Their number and origin are dictated by an \emph{endorsement policy}.
For instance, three peers may receive the transaction and simulate the chaincode independently.
The set of values in the chain, result of the execution of all valid transactions, is materialized locally in a key/value store and used as input to the simulated chaincode.
Values are associated with a version number.
The read set of a simulated transaction contains the version number of all accessed keys.
The chaincode runs isolated in a Docker container and produces a set of new values for a number of keys, forming its \emph{write set}.
Non-deterministic chaincodes produce different read and/or write sets, which can be detected by the client.
The read and write sets are compiled in an endorsement signed by the endorser and returned to the client.

The client combines the received endorsements into a \emph{transaction proposal}.
This proposal is sent to a peer, who forwards it to the ordering service.
The transaction is kept in a buffer to be eventually integrated into a new block.
Orderers do not perform any validation on transaction proposals.
A new block is proposed for consensus when its size reaches a maximal size, or after a timer expires.
Example values in Fabric~v1.2 are 99~MB and 2~seconds but must be adjusted for the target applications.

Consensus outputs chained blocks in a unique order.
A new block must be propagated to all peers to allow future transactions to operate on the new values.
Orderers send a new block to one peer in each organization.
This \emph{leader peer} is in charge of initiating the broadcast of the block to all peers in its organization.
Using a direct-send approach for this broadcast is not scalable since both the dissemination time and bandwidth usage of the leader peer increase linearly with the size of the organization.
Any organization with more than a handful of peers will, therefore, resort to gossip-based broadcast protocols.
We detail how gossip is implemented in Fabric in Section~\ref{section:gossip_in_fabric}. 

A peer receiving a new block validates all the transaction proposals it contains.
A proposal is valid if the number and origin of its endorsements satisfy the endorsement policy.
In addition, it must operate on the latest values present in the ledger.
This can be checked based on the read set of the proposal, where the version of all keys must be the same as the version in the local key/value store.
The output of all valid transactions is applied to the key/value store and the block appended to the local copy of the chain (\emph{without} re-executing the chaincodes).
Invalid transactions remain in the blockchain but have no effect besides wasted storage space.  

\subsection{Consistency of Fabric}

Even though the Fabric orderers define a total order on the submitted transactions, and peers eventually agree on the same replica of the ledger, consistency conflicts can arise in the interim, leading to invalidated transactions and lower quality of service for the clients who must prepare and resubmit new transaction proposals.
Without proper safeguards, this can cause problems at the application level such as an invalid trade in a digital marketplace.
We differentiate two types of conflicts:  \emph{proposal-time} and \emph{validation-time} conflicts.

Proposal-time conflicts can happen when two or more endorsers execute the transaction over different \emph{heights} of the ledger because they did not yet receive the same set of blocks.
Even if the chaincode is deterministic, its results can differ if one of the endorsers has not yet seen a recent write. 
This conflict can be detected at the client by comparing version numbers in the read sets, but it wastes resources and causes delays as the client must collect additional endorsements.

Validation-time conflicts can occur when two transactions issue writes to a common value without being aware of each other's execution.
This happens when two chaincodes access at least one key in common and are executed over the same content of the ledger.
These conflicting proposals may be ordered in different blocks, or at different places within the same block. 
This can cause conflicting writes characteristic of weakly consistent systems.
In Fabric the total order imposed by the ledger allows a natural \emph{earliest-writer-wins} conflict resolution policy.
The first transaction is deemed valid and its output applied to the local copy at validation, whereas the second transaction will fail the validation check. 




Validation time conflicts have a higher impact than proposal-time conflicts.
First, the client submitting a transaction only learns that it has lost the race with another transaction after its invalidation, which causes a long delay before a new transaction proposal can be prepared and resubmitted.
Second, it creates useless, invalid entries in the ledger, consuming valuable storage resources at  all peers.

Both types of conflicts occur because endorsers simulate transactions without synchronization on their state.
This is a conscious design choice led by trust assumptions.
Consistency issues are mentioned by the authors of the paper detailing Fabric~\cite{Hyperledger_EuroSys18}.
They mention that careful application design (i.e., avoiding chaincodes that write to the same value) can avoid conflicts, but it is unclear how to enforce such a strong constraint with mutually distrustful organizations.
Another suggested approach is to use conflict-free replicated data types (CRDTs)~\cite{shapiro2011conflict}, allowing concurrent writes to deterministically merge into a single value but restricting the programming model significantly. 
A third approach would be to use a \emph{lead endorser} for each chaincode  acting as a serialization point for all its executions.
This approach may avoid conflicts between instances of the same chaincode.
It is however costly and will have no impact for different chaincodes that need to access and write the same values.

\vspace{-1mm}
\section{Gossip broadcast in Fabric}
\label{section:gossip_in_fabric}

\subsection{Gossip broadcast in Fabric}

We detail in this section how epidemic/gossip-based broadcast is currently implemented in Fabric\footnote{\url{https://hyperledger-fabric.readthedocs.io/en/release-1.2/gossip.html}  provides high-level details about the objectives and implementation of gossip in Fabric.}.
Fabric uses gossip for many purposes~\cite{8645637}: peers use it to build and maintain a local view of other peers in the network; they also use it when joining the network for the first time or after a crash.
Gossip is used to disseminate metadata such as ledger height and active chaincodes.
Finally, gossip is primarily used to disseminate new blocks to all peers within an organization, who then independently validate their content and apply the results of valid transactions to their local copy of the ledger.

In this work we focus on the gossip-based dissemination of data blocks from the ordering service to the peers.
We mention that gossip is optional in this setting, and for trust reasons only allowed between peers of the same organization.
We assume, therefore, that we operate in an environment with at least one large enough organization for which a direct broadcast from the leader peer to all other peers would have a prohibitive cost.



\begin{figure}[t!]
  \centering
  \includegraphics[scale=0.75]{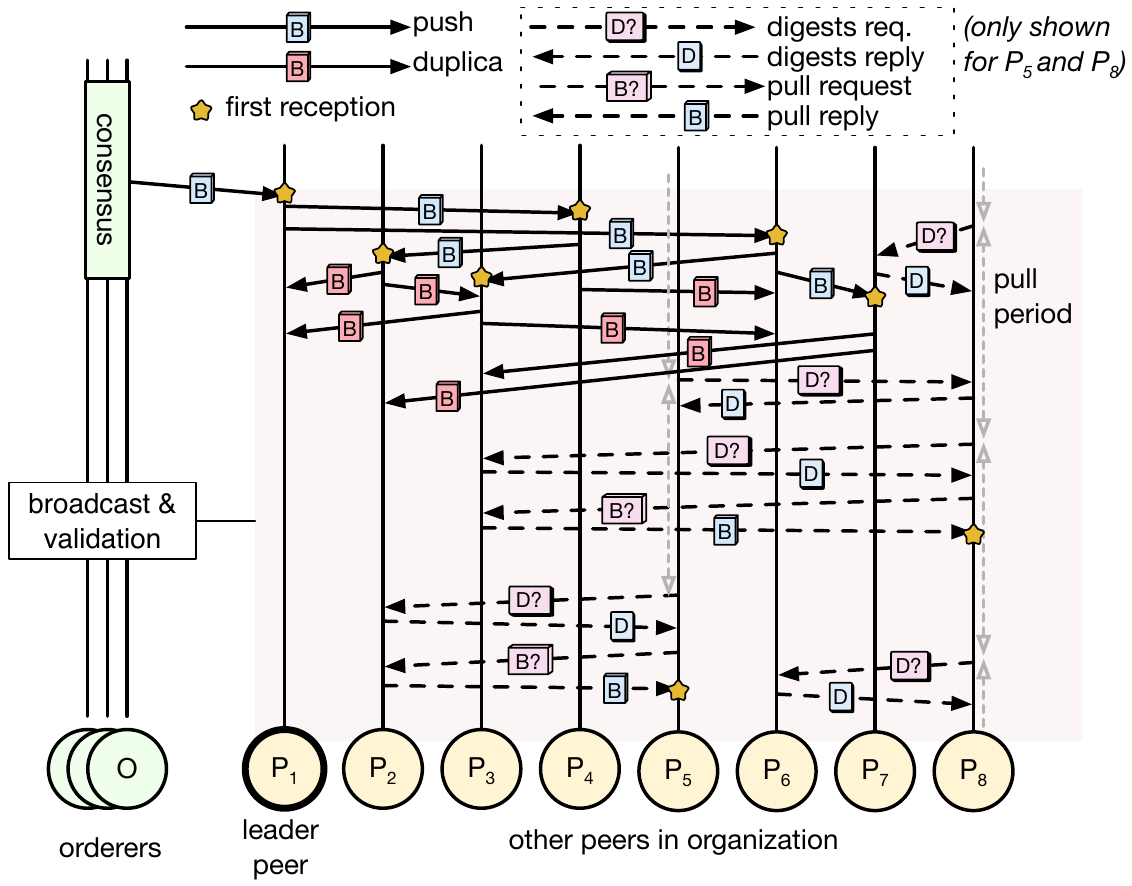}
  \caption{Dissemination of a new block in Fabric, with $\fanout=2$ and $\fanin=1$.\vspaceaftercaption}
  \label{fig:gossip_in_fabric}
  \vspace{-3mm}
\end{figure}

Figure~\ref{fig:gossip_in_fabric} illustrates the propagation of a block within an organization.
The ordering service first sends a copy of a new block to the leader peer, who initiates the gossip broadcast to the other peers in the organization. 
The gossip protocol operates on a complete graph since all peers know the identity of all other peers within the organization. Fabric combines three dissemination primitives: \emph{push}, \emph{pull} and \emph{recovery}.\footnote{Note that recovery is sometimes called \emph{anti-entropy} in Fabric documentation and in developers' communication. This creates a confusion as the term was originally used to denote \emph{pull} in early work on gossip dissemination~\cite{demers1988epidemic}. We favor the more explicit term to avoid this confusion.}

\smallskip
\noindent\textbf{Push:} Fabric uses an infect-and-die push model~\cite{EGMM2004}.
Whenever a peer receives a block for the first time (it gets infected), it pushes the block once to $\fanout$ peers chosen at random, but does not push again if it subsequently receives the same block (it pushes, then dies).
More technically, the new block is first put in a buffer, and pushed when the buffer is full or a short timer $\tpush$ expires.
Figure~\ref{fig:gossip_in_fabric} shows a push phase with $\fanout=2$. The default configuration of Fabric uses $\fanout=3$ and $\tpush=10$~milliseconds.
Some peers receive a block multiple times during the push phase (e.g. $P_1, P_2, P_3$ and $P_6$), whereas other peers do not receive it (e.g. $P_5$ and $P_8$). 

\smallskip
\noindent\textbf{Pull:}
Peers use the pull component to obtain the blocks they did not receive during the push phase. 
At every time interval~$\tpull$, every peer initiates a pull to $\fanin$ peers chosen at random.
If $\fanin=3$ and peer $P$ contacts peers $Q,R,S$, it first requests the digests of recent blocks from $Q,R,S$, which they will forward back to $P$.
From all the received digests, $P$ then requests the blocks from $Q,R,S$ that it does not already have, which are finally forwarded to $P$. 
Note that when a peer receives a new data block from a pull request, it will not directly push this block to other peers, but will only reply to pull requests.
The default parameters of the Fabric pull component are $\fanin=3$ and $\tpull=4$~seconds.
We use $\fanin=1$ in Figure~\ref{fig:gossip_in_fabric} and only show pull interactions for $P_5$ and $P_8$ in the interest of clarity.
Pull can lead to high latencies for some peers, as it is initiated infrequently.
For instance, peer P$_\text{5}$ only receives the new block on its second pull attempt, from peer~P$_\text{2}$.

\smallskip
\noindent\textbf{Recovery:} 
The recovery gossip component of Fabric allows peers to catch up with the content of the ledger by asking for a batch of missing blocks, either when they join the network, or after a long delay due to an outage, high latency or any other cause. 
At every time interval $\tantientropy$, each peer observes the metadata information containing the ledger height of the other peers in the channel. 
As opposed to the push and pull components, this is not limited to peers in the same organization. 
If a peer is behind the others, it requests the consecutive missing blocks from one of the peers with the highest ledger. 
The default parameter of the Fabric recovery component is $\tantientropy=10$~seconds.
In practice, recovery has no impact on the propagation latencies in a stable network.

\subsection{Impact of gossip on bandwidth and conflicts}

The performance of broadcast plays a direct role in the occurrence of inconsistencies in Fabric: long propagation times between orderers and peers can cause validation-time conflicts, and proposal-time conflicts are a direct consequence of uneven propagation times between orderers and peers. 
In large organizations resorting to gossip to broadcast ordered blocks, the default delays induced by the pull and recovery components (4 seconds and 10 seconds) are larger than the timer before a block is proposed for consensus (2 seconds). 
In other words, a large subset of conflicts is intrinsically caused by the broadcast \emph{tail latency}, i.e., the tail of the latency distribution corresponding to the time required for a peer to receive a new block after its addition by the ordering service.
It is therefore desirable to have a broadcast protocol that is as fast as possible in delivering new blocks to all peers for Fabric to scale without a massive increase in the occurrence of conflicts.
Similar observations have been made for eventually-consistent key-value stores in the cloud~\cite{bermbach2011eventual}.

The gossip layer also has a big impact on the bandwidth consumption of the Fabric peers. With large blocks and high throughput, the bandwidth of data blocks dominates, which can create contention and load-balancing issues, inducing additional delays and higher operating costs. 
The objective of this work is thus to analyze and optimize the broadcast layer of Fabric, with a focus on \emph{efficiency}, \emph{fairness}, and \emph{cost}.




\vspace{-1mm}
\section{Enhanced epidemic dissemination}
\label{section:efficient_gossip}

The literature on epidemic dissemination, both theoretical and practical, is vast, and gossip protocols were proposed for a wide variety of problems such as replicated database maintenance~\cite{demers1988epidemic},
propagation~\cite{Berger:2005:SVI:1070432.1070475},
membership~\cite{Jelasity2007}, 
streaming~\cite{felber2012p},
total ordering~\cite{matos2015epto},
and many distributed signal processing tasks~\cite{journals/pieee/DimakisKMRS10}.
However, there is little work that simultaneously targets practical applications supported by precise theoretical guarantees. 
On the one hand, theoretical work generally assumes that the communication takes place in synchronous rounds and that the cost of establishing the communication between peers is free, and provides asymptotic results for large deployments that disregard costly multiplicative constants
~\cite{DBLP:conf/focs/KarpSSV00}.
On the other hand, practical work is generally empirical and does not estimate the probability of imperfect dissemination.

The main advantage of the Fabric gossip module, highly practical, is its modularity: the push, pull and recovery components are fully decoupled.
This allows to push fast and to pull infrequently.

The infect-and-die push component of Fabric has two main disadvantages: its high communication overhead and its low probability of disseminating blocks to \emph{all} peers.
For instance, in our experiments with a network of $n=100$ peers and $\fanout=3$, we can easily calculate that infect-and-die push disseminates each block to an average of 94 peers with a standard deviation of 2.6, while transmitting each block in full 282 times.
The uninformed peers at the end of the push phase must resort to the infrequent pull or recovery components to receive a block, inducing a high latency.
Reaching all peers during the push phase with high probability\footnote{With high probability means with probability $O(n^{-c})$ for some $c>0$.} requires a large $\fanout$, which causes load-balancing and contention problems, and requires transmitting the block $n \cdot \fanout$ times.

We now describe the enhancements we made to the Fabric gossip module, summarized in Table~\ref{table:enhancements}.

\begin{table*}[t]
	
	\newcolumntype{C}{>{\raggedright\let\newline\\\arraybackslash\hspace{0pt}}m{0.15\linewidth} }
	\newcolumntype{D}{>{\raggedright\arraybackslash} m{0.385\linewidth} }
	\newcolumntype{E}{>{\raggedright\arraybackslash} m{0.385\linewidth} }
	
  \centering
  \begin{tabular}{CDE}
	\toprule
	\textbf{Enhancement} & \textbf{Description} & \textbf{Benefit} \\
    \midrule
    \emph{Infect-upon-contagion} push dissemination & Peers forward blocks at every round during which they receive them. & Better load balancing and better control over the number of outgoing messages for each peer. \\ \hline
    \emph{Digests} for the push phase & Peers forward a digest of the blocks they receive instead of the entire block. & Remove communication overhead caused by receiving the same full block multiple times.\\ \hline
    \emph{Randomization} of the initial gossiper & Leader peers initiate gossip by sending the block to an initial peer, who initiates the dissemination to $\fanoutleader$ peers. & Reduce the burden on leader peers and increase fairness.\\ \hline
    \emph{Removal} of the pull component & Deactivate the pull component.  & Avoid overhead caused by the now-unnecessary pull component. \\
    \bottomrule
  \end{tabular}
  \vspace{+1mm}
  \caption{Summary of the enhancements to the Fabric gossip layer.}
  \vspace{-6mm}
\label{table:enhancements}
\end{table*}



\smallskip
\noindent\textbf{Infect-upon-contagion push dissemination:}
We replace the infect-and-die push component by an \emph{infect-upon-contagion} push algorithm~\cite{koldehofe2004simple}.
In this model, peers forward blocks at every round during which they receive them.
The main advantage of the infect-upon-contagion push over the well-known infect-forever model~\cite{eugster2004epidemic} is its load balancing, as it does not impose an undue burden onto the peer initiating the rumor.
Our protocol requires a stopping condition: we attach a counter $r$ to each block, initialized at 0.
When a peer receives a block~$b$ with counter $r=k$ (the exact pair $(b,k)$) for the first time, it increments its counter to $k+1$ and forwards $b$ to a sample of $\fanout$ other peers chosen uniformly at random.
The dissemination stops when the counters of the blocks being disseminated locally reach an agreed-upon time-to-live $\TTL$ value. Note that the notion of rounds, even asynchronous, is unnecessary in this setting.

The probability of imperfect dissemination $p_e$, i.e., the probability that a block does not reach all peers during the push phase, depends on $n$, $\fanout$ and $\TTL$.
We can calculate $p_e$ very precisely.
Our analysis is summarized in the appendix.
We target $p_e = 10^{-6}$ for a network of $n=100$ peers in our experiments. 
We consider two configurations achieving this goal: (1) $\fanout=\lfloor\ln n \rfloor = 4$ and $\TTL=9$; and (2) $\fanout=2$ and $TTL=19$.
If $p_e=10^{-6}$ is insufficient, increasing $\TTL$ 
from 9 to 12 with $\fanout=4$ leads to $p_e=10^{-12}$, which is in all likelihood smaller than the probability of hardware crash for a given peer.
Despite these seemingly high $\TTL$ values, the absence of predetermined rounds makes this phase significantly faster than the frequency of the pull and recovery components.
Note also that $\TTL$ varies slowly with $n$; we can, therefore, store a small number of $\TTL$ values for $(n,p_e)$ pairs in a lookup table.
Peers can adjust $\TTL$ using the lowest upper bound for the number of peers appearing in the table. 

We also remove the $\tpush=10$ms timer for data blocks embedded in Fabric, not to improve the latency but to ensure unbiased randomness.
More precisely, the timer causes bias because peers receiving a block $b$ with different counter values during a $10$ms interval put them in the same buffer and transmit them to the same $\fanout$ peers, reducing the number of messages, which increases the probability of imperfect dissemination $p_e$ above the theoretical guarantees.
The simplest way to solve this problem is to set $\tpush=0$ for data blocks, ensuring that each pair $(b,k)$ is forwarded to a random sample of $\fanout$ peers. 

\smallskip
\noindent\textbf{Digests for the push phase:}
As we increase $TTL$, at the end of the dissemination, most if not all peers are already informed and keep sending the same blocks to each other.
This results in an unacceptable communication overhead.
We thus introduce digests for the push phase, as done for many existing push protocols.
When a peer receives a block whose counter $r$ is below $TTL$, it first selects a sample of $\fanout$ peers at random and sends them a digest of the block.
The peers who do not already have the block will request it, and the block will then be pushed to them. 
We can further improve this by pushing without digest until $\TTLdirect$ while there are little collisions, and push with a digest until $TTL$.
With $n=100$ and $\fanout = 4$, we can set $\TTLdirect=2$.
In our experiments on a LAN, this feature has shown little effect on latency and can be omitted.

The number of messages per block sent by our push protocol is $k \ln n$ where $k$ is a large constant.
This is unavoidable to reach all peers with high probability without using pull.
With a digest, we ensure that large blocks are only transmitted $n+o(n)$ times, with the other messages being small digests.
With push-pull algorithms, $\Theta(n \ln n \ln n)$ messages are optimal~\cite{DBLP:conf/focs/KarpSSV00}, but this assumes that peers know when blocks arrive in the network, and that establishing the communication between peers is free. 
Without these unrealistic and impractical assumptions, a large time between rounds must be respected.
Our protocol is, therefore, orders of magnitude faster than this variant. 

\smallskip
\noindent\textbf{Randomization of the initial gossiper:}
In Fabric, the leader peer, who receives all the blocks from the ordering service, always initiates the dissemination.
Its downstream bandwidth is thus $\fanout$ times higher than that of regular peers.
To alleviate this burden, we fix $\fanoutleader=1$.
On expectation, this splits the beginning of the gossip phase evenly among the other network peers.

\smallskip
\noindent\textbf{Removal of the pull component:}
By setting the probability of imperfect dissemination low enough, the pull component loses its usefulness, thus we get rid of it.
In the unlikely occurrence that a peer fails to receive a block at the end of the push phase, it will be fetched during recovery.
The recovery component serves other fundamental needs such as recovery after crashes and we keep it as it is.

\vspace{-1mm}
\section{Evaluation}
\label{section:evaluation}

We evaluate our improved Fabric gossip module and compare it to the original version.
Our evaluation focuses on three criteria: latency of block dissemination, the impact of latency on consistency conflicts, and bandwidth consumption.

\subsection{Experiment setup}

We use version 1.2 of Fabric.
We set up a network of 100 peers belonging to a single organization, one client, and a CFT ordering service consisting of four Kafka~\cite{kafka} nodes and three Zookeeper~\cite{hunt2010zookeeper} nodes, corresponding to the default configuration for a Kafka-based setup.
All nodes are deployed on a cluster of 15 servers equipped with 8-core L5420 Intel\textregistered{} Xeon\textregistered{} CPUs at 2.5~GHz and 8~GB of RAM, interconnected using 1~Gbps Ethernet.
All components run inside Docker containers on at least one dedicated core.

In our first set of experiments, we use a chaincode from the Fabric high-throughput network example that models n cryptocurrency asset whose value is frequently modified~\cite{high-throughput}.
We use the following configuration for orderers: blocks contain at most 50~transactions, with a timeout of 2 seconds if less than 50 transactions are received.
During every test, we sequentially initiate 50,000 transactions so that a full block containing 50~transactions is created every $\approx 1.5$ second. 
This results in 1,000 blocks of $\approx 160$~KB per block.


\subsection{Evaluation baseline (original Fabric gossip module)}

For comparison purposes, we initially test the original gossip module of Fabric, without any modification, using the experimental setup described above.
We use the default parameters for the push, pull and recovery components: the push component uses $\fanout=3$ and a timer of $\tpush=10$~ms, the pull component uses $\fanin=3$ and is executed by each peer every $\tpull=4$~seconds, and peers execute the recovery component every $\tantientropy = 10$~seconds. 

\begin{figure}
  \centering
\includegraphics[width=1\columnwidth]{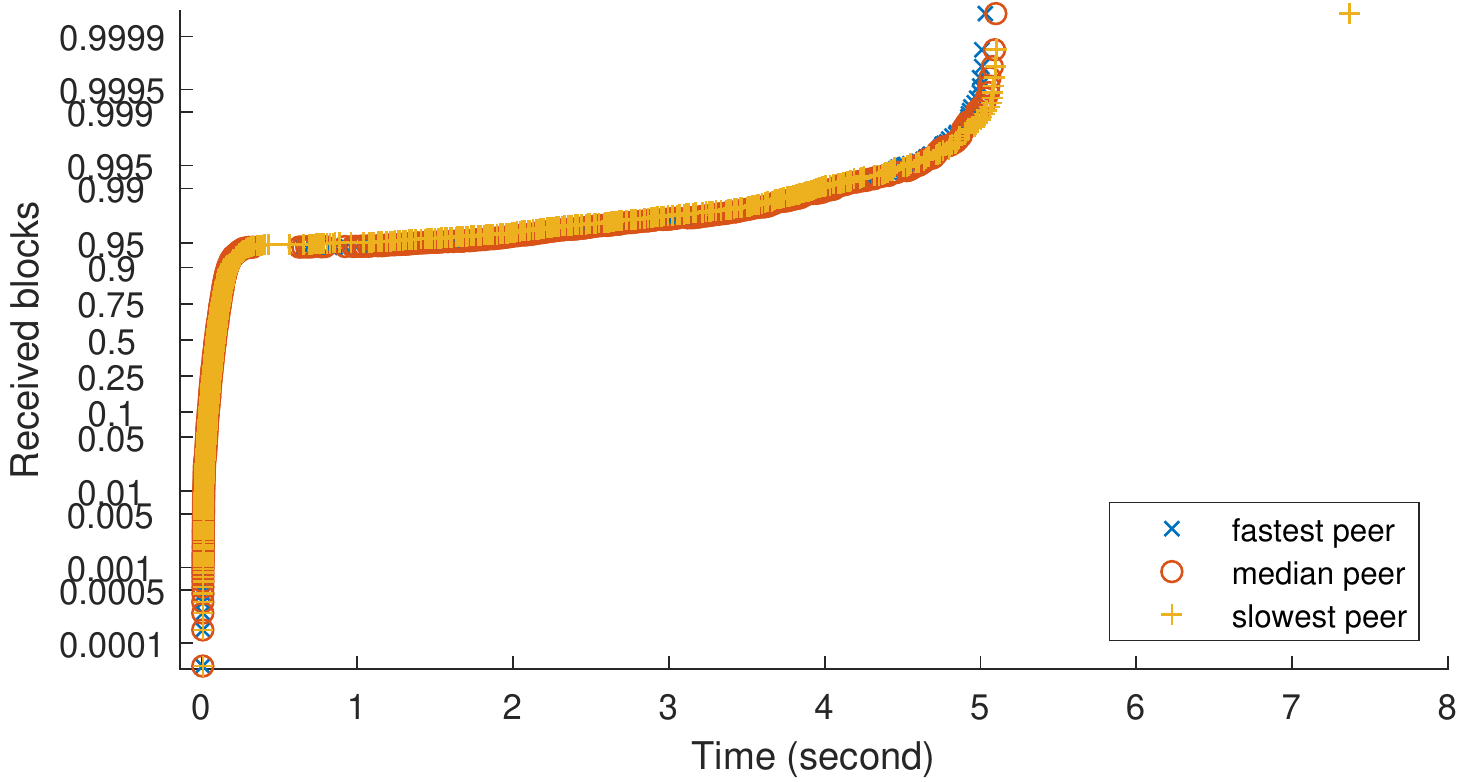}
\vspacebeforecaption
\caption{Latency at the peer level using the original gossip module of Fabric.
\vspaceaftercaption}
\label{fig:baseline-peer}
\end{figure}

\begin{figure}
  \centering
\includegraphics[width=\columnwidth]{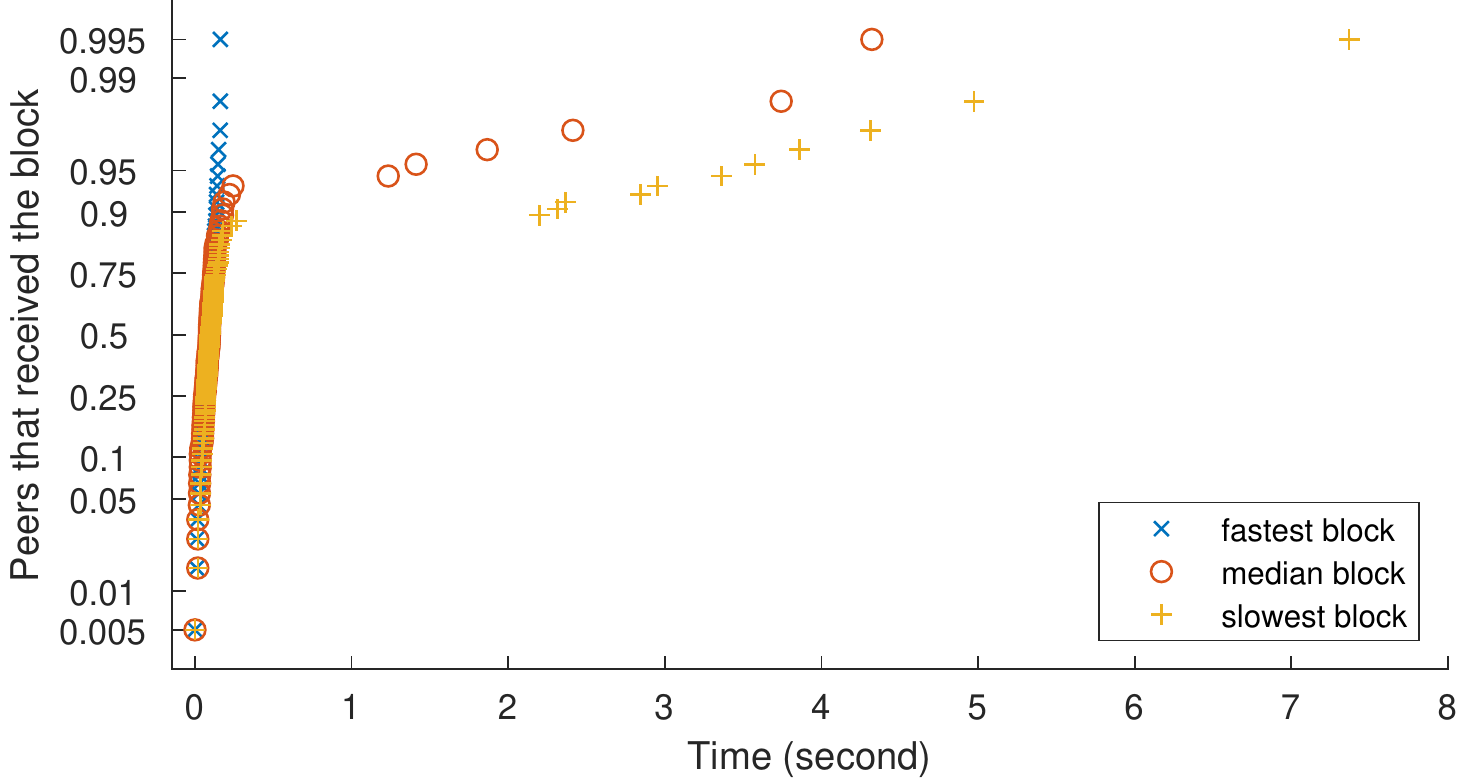}
\vspacebeforecaption
\caption{Latency at the block level using the original gossip module of Fabric.\vspaceaftercaption}
\label{fig:baseline-block}
\end{figure}

Figure~\ref{fig:baseline-peer} first illustrates the latency at the peer level by showing how much time peers take to receive blocks starting from the beginning of their dissemination (i.e. their reception by the contact peer from the orderer nodes).
It includes three CDFs for peers with the slowest, median and fastest average latencies.
Figure~\ref{fig:baseline-block} then illustrates the latency at the block level by showing how fast given blocks reach each network peer.
We include three CDFs for the slowest, median and fastest disseminated blocks. 
Figures~\ref{fig:baseline-peer} and~\ref{fig:baseline-block} are probability plots with a logarithmic scale based on a logistic distribution, for two reasons.
First, this allows focusing on tail latencies.
Second, most push dissemination protocols grow like a logistic function (i.e., exponentially at the beginning of the dissemination, followed by slower growth to inform the last uninformed processes), as we observe by the linear fit on the left side of the figures.
The fat and long tail is obvious in both figures, and corresponds to the transition from the fast push phase, to the much slower pull phase.

\begin{figure}
  \centering
\includegraphics[width=\columnwidth]{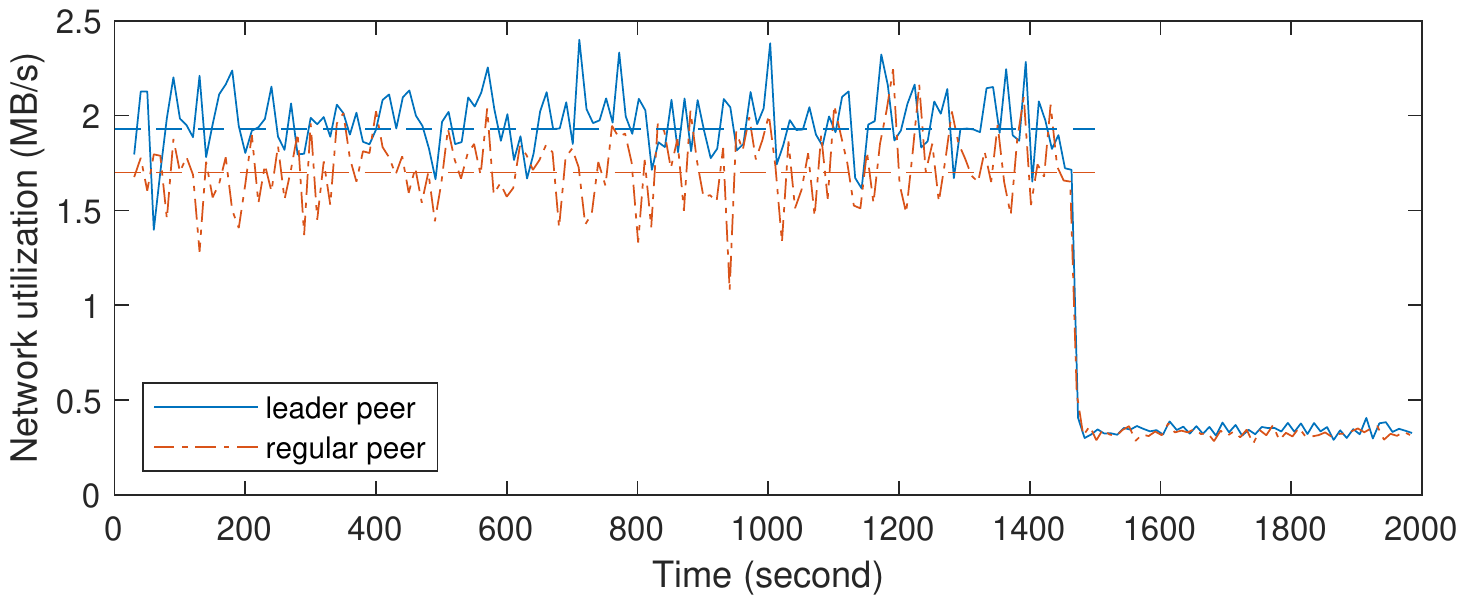}
\vspacebeforecaption
\caption{Bandwidth consumption for the leader peer and a regular peer using the original gossip module of Fabric. For readability purposes the bandwidth is aggregated at intervals of 10 seconds. Dotted lines show the average bandwidth.
\vspaceaftercaption}
\label{fig:baseline-bandwidth}
\end{figure}

Figure~\ref{fig:baseline-bandwidth} shows the bandwidth consumption for the leader peer and another peer chosen at random (all the non-leader peers behave similarly).
Note that for readability purposes we aggregated the bandwidth for each interval of 10 seconds, thus the highest traffic spikes of $\approx 4$~MB/s do not appear on the figure. 
The generation of transactions ends after $\approx$1,500~seconds.
The figure also shows the 0.4~MB/s background traffic of all the tasks when the network is in an idle state from 1,500 to 2,000 seconds, illustrating the relevance of optimizing the bandwidth requirements of the epidemic dissemination.
The main difference between the leader peer and the other peers is that the leader peer receives every block from the ordering service once and disseminates it to $\fanout$ other peers, whereas the other peers do not always do this because the infect-and-die push phase for some of the blocks terminates without reaching them.

\subsection{Evaluation of our enhanced Fabric gossip module}

We now test the performance of our enhanced Fabric gossip module and compare it to the original baseline. 
We select the parameters to obtain a probability of imperfect dissemination of $10^{-6}$ in an organization of $n=100$ peers, that is, the probability that a block does not reach all the peers during the push phase and requires the recovery component is at most $10^{-6}$. As discussed in Section~\ref{section:efficient_gossip}, we emphasize once more that $\TTL$ can be calculated precisely for any $n$, and since it varies slowly with $n$ a few values can be stored in a lookup table. 
We first set $\fanoutleader=1$ to decrease the bandwidth burden on the leader peer.
The peer who receives the block from the leader peer initiates the gossip.
We also remove the pull phase entirely due to its redundancy with the recovery component.
We finally set $\tpush=0$ for data blocks.  
Surprisingly, removing this counter has a detrimental effect on the average end-to-end latency, however as discussed in Section~\ref{section:efficient_gossip} this is the simplest way to ensure unbiased randomness, leading to the theoretical dissemination guarantees of our protocol, without further modification to the Fabric codebase.

We note that due to the chosen parameters, none of our experiments had to resort to the recovery component at any point, besides when it was triggered in the middle of a push phase. 
It was, for dissemination purposes, never required, but would have been at some point had we experimented with millions of blocks. 

\begin{figure}
	\centering
	\includegraphics[width=\columnwidth]{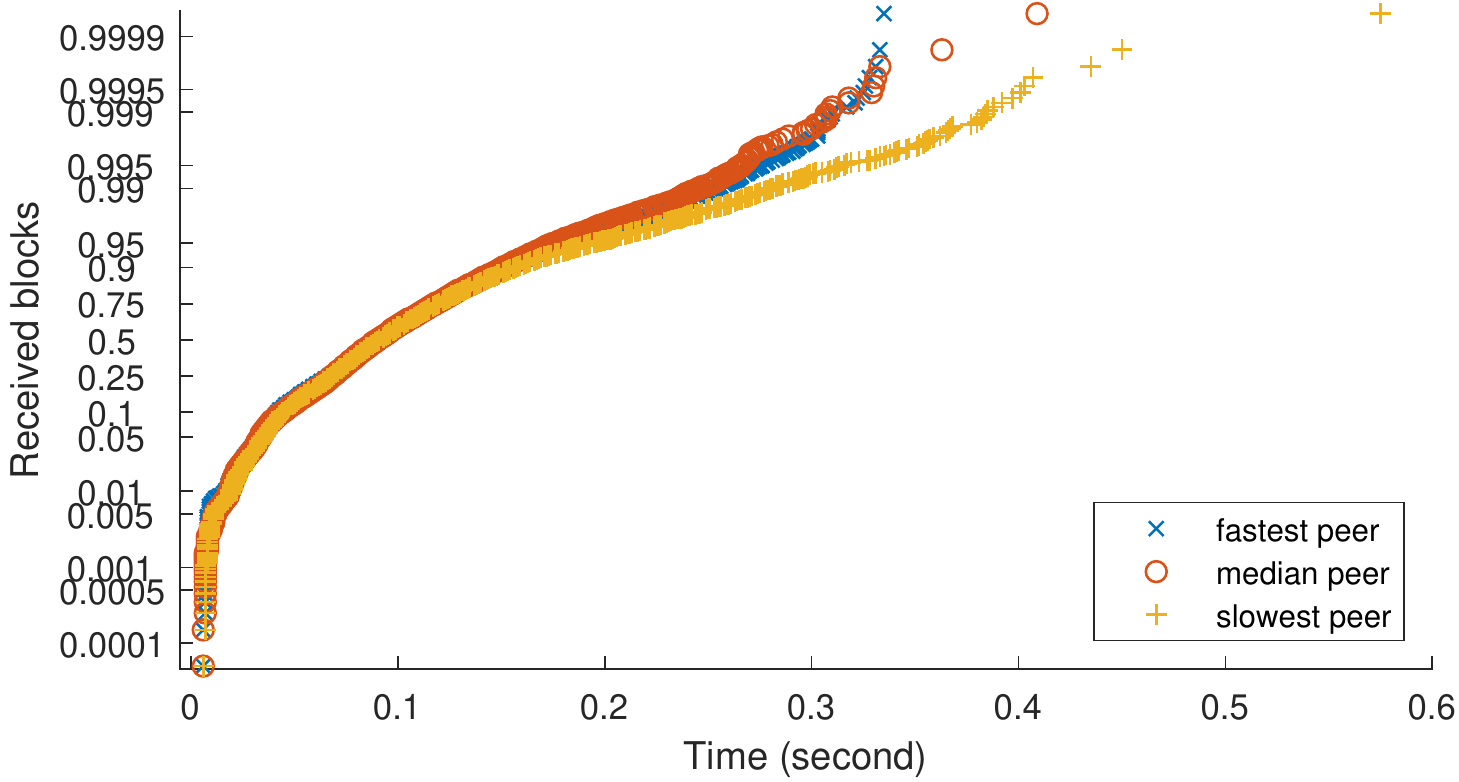}
	\vspacebeforecaption
	\caption{Latency at the peer level using the enhanced gossip module of Fabric with $\fanout=4$ and $\TTLdigest=9$.
	\vspaceaftercaption}
	\label{fig:fanout-4-peer}
\end{figure}

\begin{figure}
	\centering
	\includegraphics[width=\columnwidth]{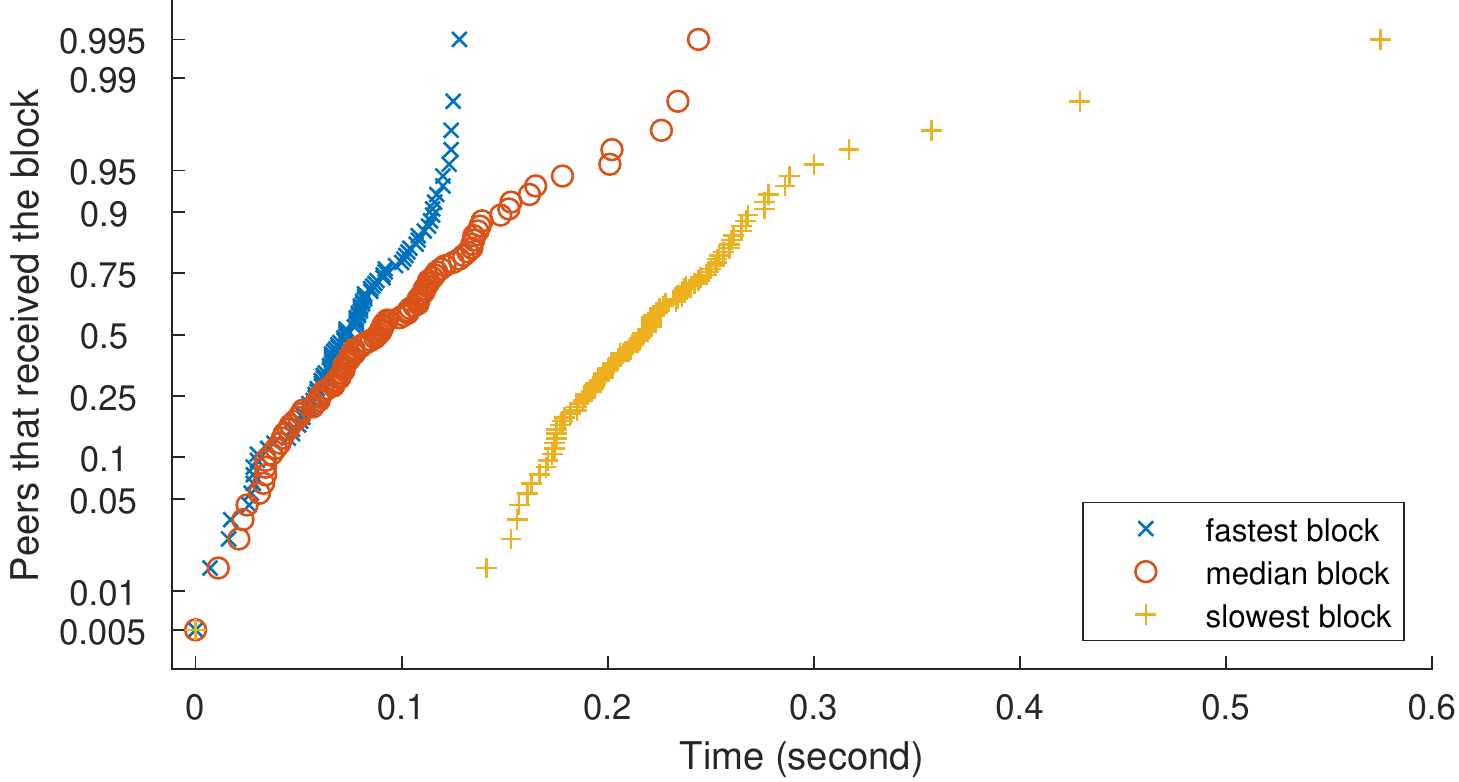}
	\vspacebeforecaption
	\caption{Latency at the block level using the enhanced gossip module of Fabric with $\fanout=4$ and $\TTLdigest=9$.\vspaceaftercaption}
	\label{fig:fanout-4-block}
\end{figure}

\begin{figure}
	\centering
	\includegraphics[width=\columnwidth]{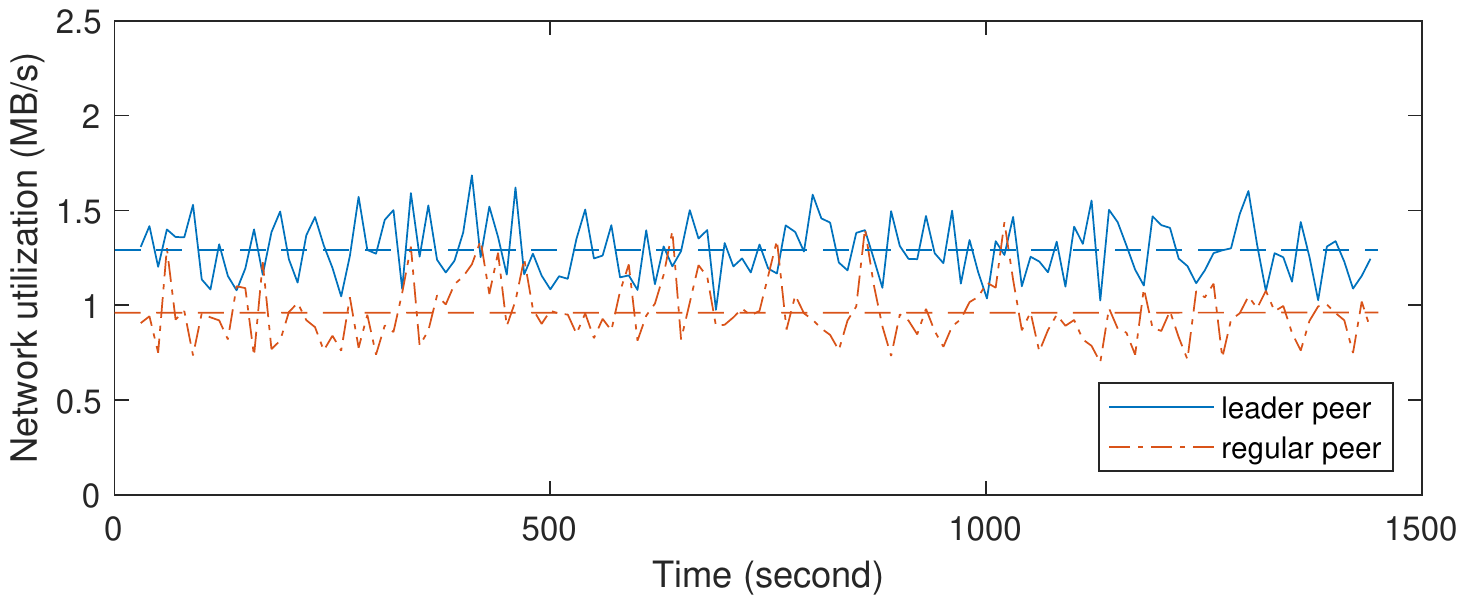}
	\vspacebeforecaption
	\caption{Bandwidth consumption for the leader peer and a regular peer using the enhanced gossip module of Fabric with $\fanout=4$ and $\TTLdigest=9$. For readability purposes the bandwidth is aggregated at intervals of 10 seconds. Dotted lines show the average bandwidth.
	}
	\label{fig:fanout-4-bandwidth}
\end{figure}

In a first evaluation, we set $\fanout = \lfloor \log n \rfloor = 4$ and $\TTLdigest=9$.
We further set $\TTLdirect=2$ because collisions are rare enough in the first two rounds to avoid sending digests.
As done for the baseline, Figures \ref{fig:fanout-4-peer}, \ref{fig:fanout-4-block} and~\ref{fig:fanout-4-bandwidth} respectively show the latency at the peer level, the latency at the block level, and the bandwidth consumption for the leader peer and other peers.
While the original configuration of Fabric requires between one and six seconds to reach the last 5\% of the peers, the enhanced gossip module disseminates all the blocks to all the peers in less than half a second.
Note that the curves in Figures \ref{fig:fanout-4-peer} and~\ref{fig:fanout-4-block} are almost linear, which we expect from probability plots with a logarithmic scale based on a logistic distribution.
The gentler slope to reach the last $5\%$ of the peers/blocks is not due to the theoretically expected propagation of the epidemic dissemination but to other sources of delay.
For example, the yellow curve on the right side of Figure \ref{fig:fanout-4-block} corresponds to the slowest block in this experiment.
We observe that this block took $\approx 0.15$~second to reach the peer initiating the broadcast, due to a delay at the leader peer.

As for the bandwidth, we observe comparing Figures \ref{fig:baseline-bandwidth} and \ref{fig:fanout-4-bandwidth} that the enhanced module decreases the bandwidth consumption of regular peers by more than $40\%$; the total network bandwidth consumption also decreases by $40\%$.
We emphasize that the difference will be even more significant with larger blocks, since their dissemination dwarfs the background bandwidth of all the other system tasks.

\begin{figure}
  \centering
  \includegraphics[width=\columnwidth]{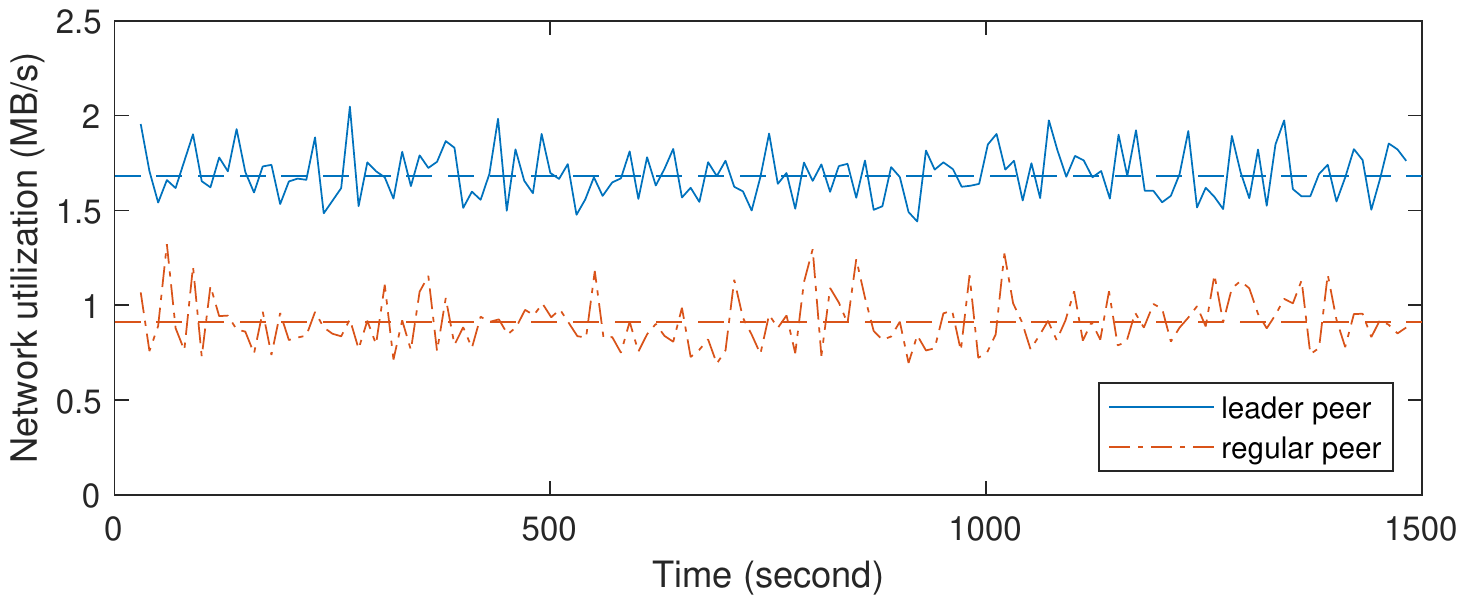}
  \vspacebeforecaption
  \caption{Bandwidth consumption for the leader peer and a regular peer using the enhanced gossip module of Fabric with $\fanoutleader=\fanout=4$ and $\TTLdigest=9$. For readability purposes the bandwidth is aggregated at intervals of 10 seconds. Dotted lines show the average bandwidth.
  \vspaceaftercaption} 
\label{fig:fanout-4-bandwidth-no-proxy}
\vspace{-2mm}
\end{figure}

\begin{figure}
  \centering
  \includegraphics[width=\columnwidth]{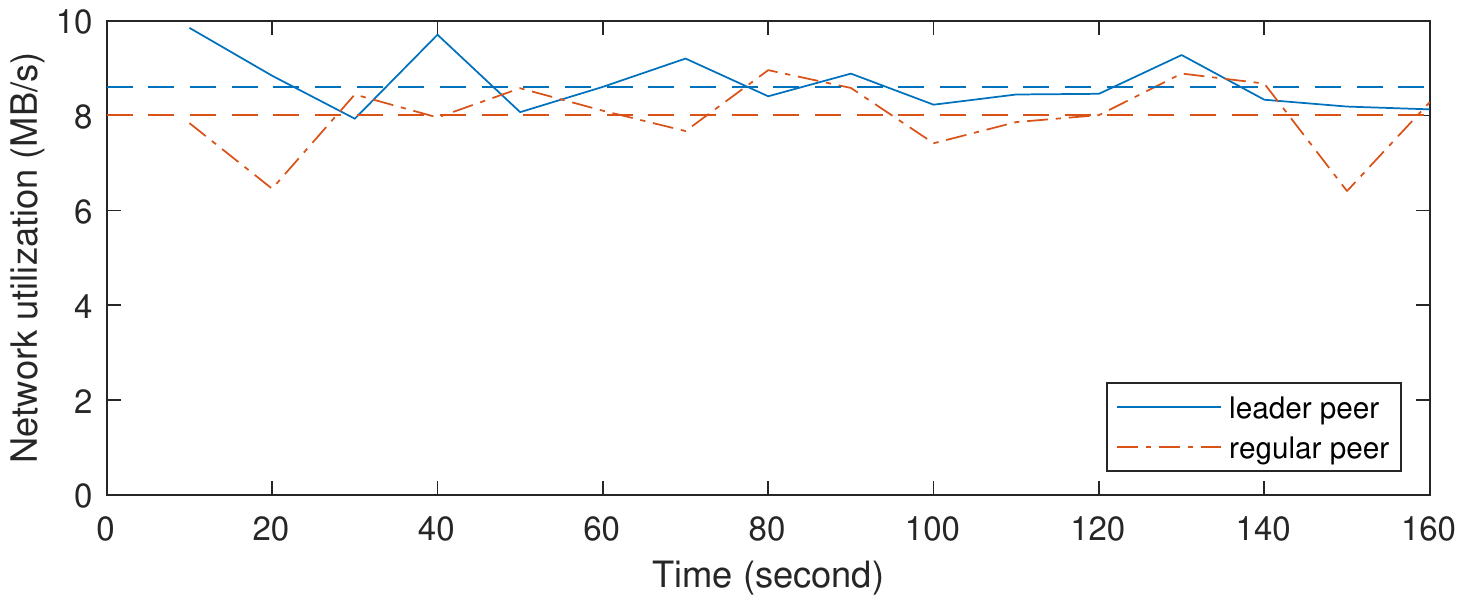}
  \vspacebeforecaption
  \caption{Bandwidth consumption for the leader peer and a regular peer using the enhanced gossip module of Fabric, but without sending digests before pushing blocks to other peers with $\fanout=4$ and $\TTLdigest=9$. For readability purposes the bandwidth is aggregated at intervals of 10 seconds. Dotted lines show the average bandwidth.
  \vspaceaftercaption} 
\label{fig:fanout-4-bandwidth-no-digest}
\vspace{-3mm}
\end{figure}

Note that setting $\fanoutleader=1$ for the leader peer and adding digests are not useless luxuries. 
Figure~\ref{fig:fanout-4-bandwidth-no-proxy} reports the bandwidth usage of the same evaluation, with the exception that the leader peer uses $\fanoutleader=\fanout=4$ like the other nodes.
We observe that the bandwidth consumption of the leader peer in this scenario is much higher than that of other peers, which could cause contention issues with large blocks and/or high-throughput applications. This is easy to explain: on expectation, random peers receive and transmit each block once, whereas the leader peer propagates every block to $\fanout$ peers. It is preferable to delegate this initial broadcast phase uniformly at random among the 99 other peers. 
Worse, Figure~\ref{fig:fanout-4-bandwidth-no-digest} reports the bandwidth of the same evaluation when peers systematically push blocks without digests, which jumps to $8$~MB/s.
Once more than $\frac{n}{\log n}$ peers are informed after the first three rounds, the number of collisions increases quickly to reach the point that peers already informed keep exchanging the same blocks between each other. 

\begin{figure}
  \centering
  \includegraphics[width=1\columnwidth]{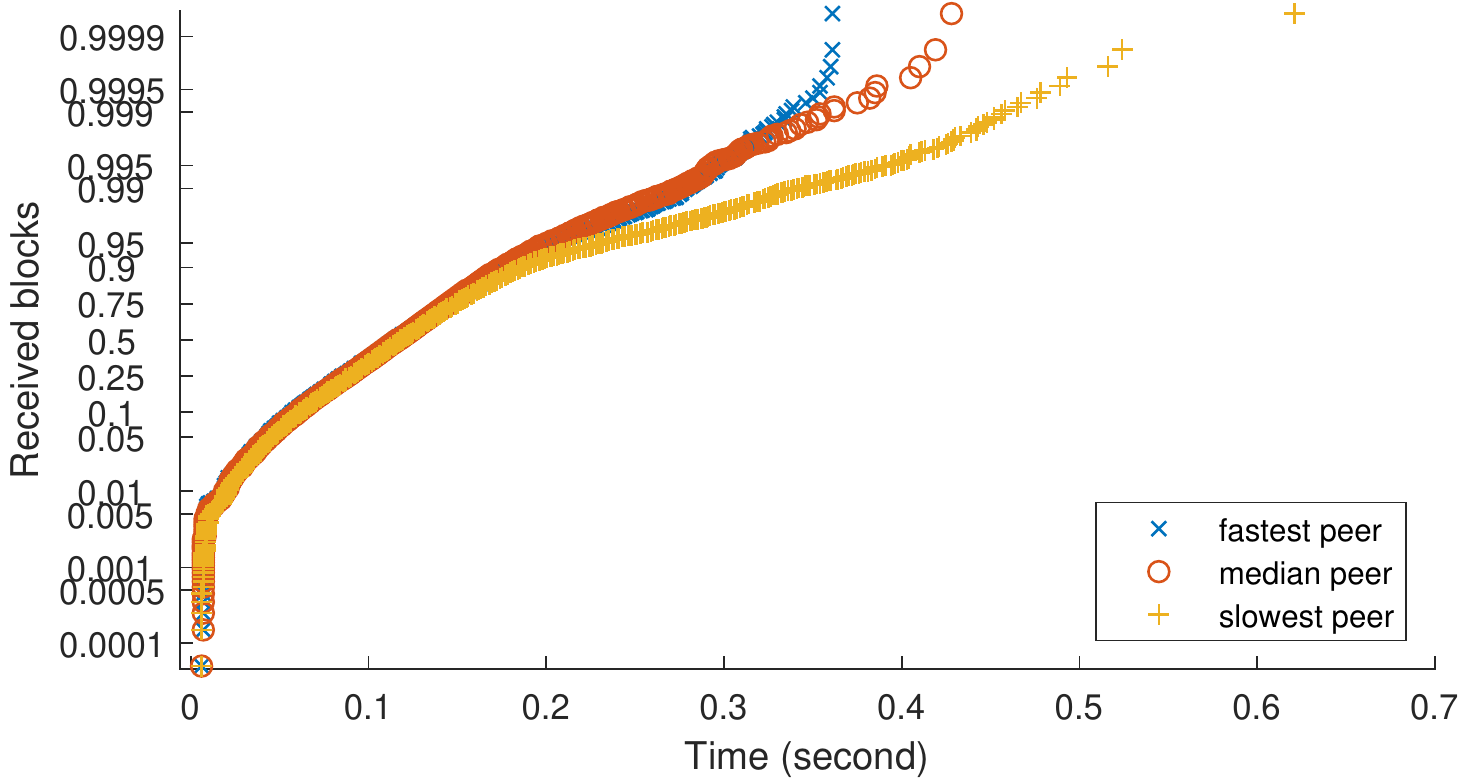}
  \vspacebeforecaption
  \caption{Latency at the peer level using the enhanced gossip module of Fabric with $\fanout=2$ and $\TTLdigest=19$.\vspaceaftercaption}
\label{fig:fanout-2-peer}
\vspace{-2mm}
\end{figure}

\begin{figure}
  \centering
  \includegraphics[width=1\columnwidth]{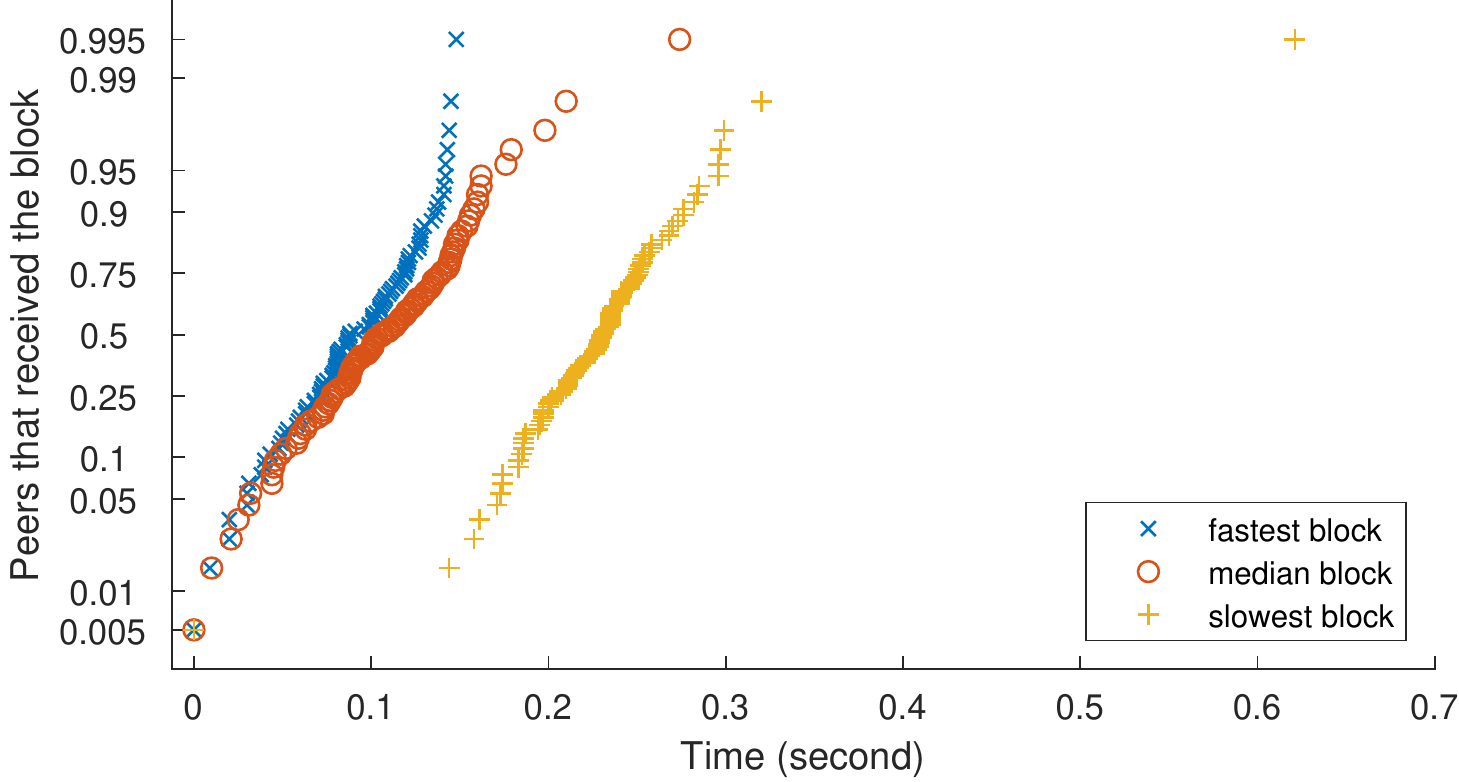}
\vspacebeforecaption
\caption{Latency at the block level using the enhanced gossip module of Fabric with $\fanout=2$ and $\TTLdigest=19$.\vspaceaftercaption}
\label{fig:fanout-2-block}
\vspace{-2mm}
\end{figure}

\begin{figure}
  \centering
  \includegraphics[width=1\columnwidth]{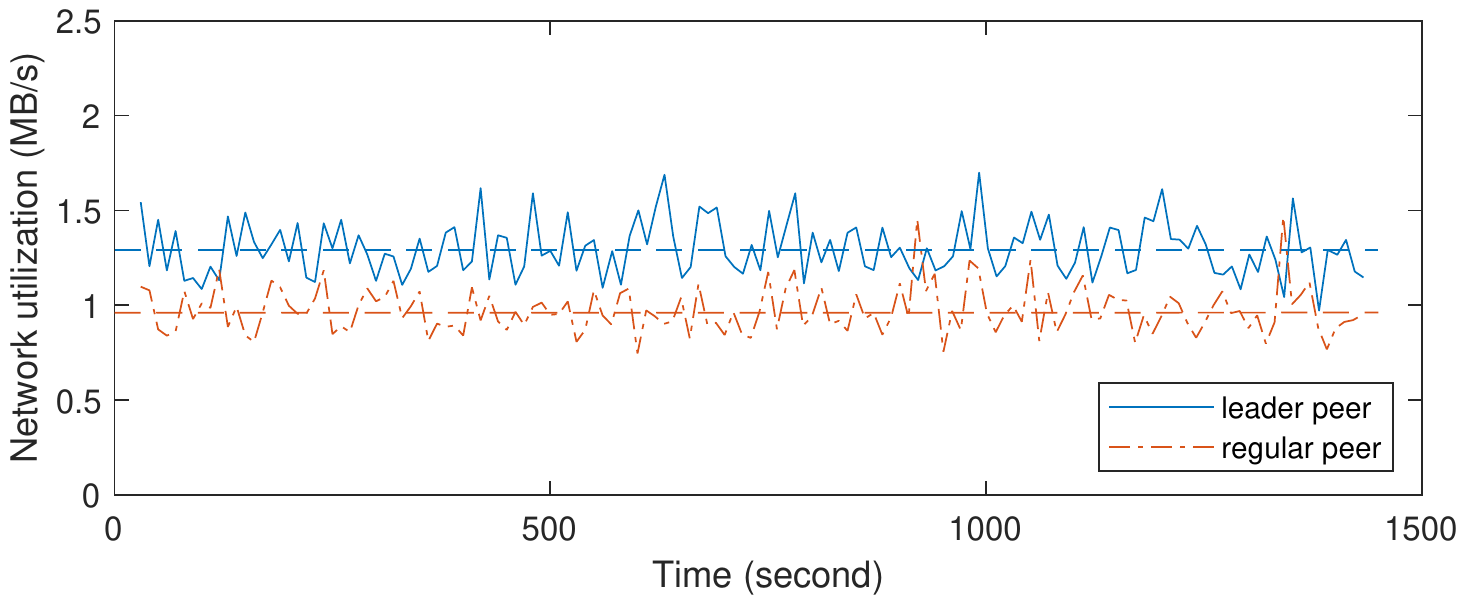}
  \vspacebeforecaption
  \caption{Bandwidth consumption for the leader peer and a regular peer using the enhanced gossip module of Fabric with $\fanout=2$ and $\TTLdigest=19$. For readability purposes the bandwidth is aggregated at intervals of 10 seconds. Dotted lines show the average bandwidth.
  \vspaceaftercaption}
\label{fig:fanout-2-bandwidth}
\vspace{-3mm}
\end{figure}

In a second evaluation, we set $\fanout = 2$ and $\TTLdigest=19$, which also guarantees that a block is disseminated to every peer during the push phase with probability at least $1-10^{-6}$. We can use $\TTLdirect=3$ in this setting.
Once again, Figures \ref{fig:fanout-2-peer}, \ref{fig:fanout-2-block} and \ref{fig:fanout-2-bandwidth} respectively show the latency at the peer level, the latency at the block level, and the bandwidth consumption for the leader peer and other peers.
We observe by comparing Figures \ref{fig:fanout-4-bandwidth} and \ref{fig:fanout-2-bandwidth} that the average and overall bandwidth consumption remains essentially unchanged. This is not surprising since the total number of transmitted digests depends on the probability of imperfect dissemination, fixed at $10^{-6}$ in both evaluations.
We also observe that decreasing $\fanout$ from 4 to 2 halves the slope of the curves in Figure~\ref{fig:fanout-2-peer} compared to Figure~\ref{fig:fanout-4-peer}, which is again expected. The interesting observation, however, is that the tails and worst-case latencies of both experiments are rather similar.
This means that $\fanout=4$ is an aggressive choice: while it speeds up the early stages of the dissemination, it stops being the dominant source of delay for the slowest blocks and peers.
With a more uniform load balancing, $\fanout=2$ is a good choice for our network.

\subsection{Impact on Fabric consistency conflicts}

We finally evaluate the impact of broadcast performance on the occurrence of consistency conflicts.
We focus on validation-time conflicts and therefore use a single endorsing peer.
We vary the timer for the generation of new blocks, while issuing a fixed rate of five transactions per second.
New blocks are only used by peers after their validation, which takes a time proportional to the number of transactions per block, about 50~ms per transaction in our evaluation.
%
Small block generation periods and therefore smaller blocks validation times put more emphasis on the impact of dissemination latency on the occurrence of conflicts.
We compare the amount of consistency errors between the original Fabric gossip module, and our enhanced gossip module using $\fanout=4$ and $\TTLdigest=9$.


\begin{table}
  \centering
  \begin{tabular}{c c c c c c}
	\toprule
	  \textbf{Block} & & \textbf{Validation} & \multicolumn{3}{c}{\textbf{Conflicts w/ different gossip}} \\ 
	  \textbf{period} & \textbf{Tx/block} & \textbf{time} & Original & Enhanced & Difference \\
	\midrule
	  2~s & 10 & 0.5~s & 803 & 664 & -17\% \\
	  1.5~s & 8-9 & 0.37~s & 814 & 653 & -20\% \\
	  1~s & 5-6 & 0.25~s & 763 & 564 & -26\% \\
	  0.75~s & 4-5 & 0.19~s & 823 & 527 & -36\% \\
    \bottomrule
  \end{tabular}
  \vspace{+1mm}
  \caption{Invalidated transactions under different block periods. Each value is the average over 
  five experiments.}
\label{table:invalidated_transactions} \vspace{-2mm} \vspace{-2mm}
\end{table}

We set up an experiment with a simple chaincode that increments one of 100 integer values initialized to 0, and stored in the ledger.
Each integer is incremented 100 times. 
We use a random permutation of the order of increments between each round of increments.
This results in a total of 10,000 transactions.
Incrementing an integer requires reading its current value.
Two increments based on the same base value will lead to a validation-time conflict.
We do not resend conflicted transactions; the difference between 10,000 and the sum over all counters in the final ledger gives the number of validation-time conflicts.
Table \ref{table:invalidated_transactions} presents our results for timer values varying from 0.75~s to 2~s.
Each entry in the table is the average number of conflicts over five experiments. 

The number of conflicts when using the original gossip module ranges from 763 to 823 on average.
Somehow surprisingly, we observe that the number of conflicts is essentially stable with the  generation period.
The reason is that while broadcast takes a few hundred milliseconds to disseminate a block to 95\% of the peers, it takes several seconds to reach the slowest 5\% peers (Figure~\ref{fig:baseline-block}). 
This tail dissemination latency dominates the ordering and validation times, no matter what block period we use.

Our enhanced gossip module, with its small average and worst-case dissemination delays (Figure~\ref{fig:fanout-4-block}), significantly improves upon the original Fabric gossip.
As shown in Table~\ref{table:invalidated_transactions}, the average number of conflicts decreases with the block period, to 664 with a period of 2~s, and down to 527 for a block period of 0.75~s.
This corresponds to reductions in the number of invalidated transactions varying from 17\% to 36\% compared to the original Fabric implementation.

\vspace{-1mm}
\section{Related work}
\label{section:related}

Although the analysis and optimization of the native Fabric implementation is getting a lot of attention, little of it focuses on the gossip layer or block propagation.
The only work we are aware of in this direction is a poster presentation by Barger \emph{et~al.}~\cite{barger2017scalable}, who discuss early work on the BlockStorm epidemic protocol for Fabric.
The objective of BlockStorm is to resist the impact of adversarial peers during the broadcast.
FairLedger, proposed by Lev-Ari~\emph{et~al.}~\cite{fairledger2019opodis}, introduces a permissioned BFT blockchain protocol tested on Hyperledger Iroha, an alternative framework to Fabric, targeting fairness in terms of opportunity for participants to append their transactions to the ledger. 
The core of the protocol is again focused on an adversarial-resistant broadcast, able to detect peer misbehavior such as withholding messages. 
These solutions represent a complementary line of work that we intend to explore if Fabric extends gossip to work across organizations.

The throughput and latency of Fabric were evaluated in multiple studies.
Sukhwani \emph{et~al.}~\cite{sukhwani2018performance} model the performance of Fabric using stochastic Petri nets.
They show that larger blocks improve throughput by reducing the bottleneck at the ordering service, at the price of higher latency.
Baliga \emph{et~al.}~\cite{baliga2018performance} propose micro-benchmarks to measure the transaction throughput and latency of Fabric in various scenarios.
Thakkar \emph{et~al.}~\cite{thakkar2018performance} propose optimization to the \emph{endorsement} phase using caching and parallelization, while Javaid \emph{et~al}~\cite{javaid2019optimizing} propose optimizations to the \emph{validation} phase of Fabric.
Kwon and Yu~\cite{kwon2019performance} also evaluate the performance of the ordering and endorsement phases in Fabric.
None of the work referred in this paragraph was tested with a large enough number of peers for broadcast gossip to play a significant role on fairness or throughput.

Other works compare the performance of different versions of Fabric, and compare Fabric with other blockchain frameworks.
A comparison between Fabric (v0.6 and v1.0) and Ripple~\cite{schwartz2014ripple} is presented by Han \emph{et~al.}~\cite{han2018evaluating}.
It focuses on the impact of the BFT consensus.
The authors' conclusion is that Byzantine consensus does not scale well for large blockchain deployments, although Fabric reaches a better throughput than Ripple.
A limited comparison between Fabric and a private deployment of Ethereum using a single blockchain node is proposed by Pongnumkul \emph{et~al.}~\cite{pongnumkul2017performance}.
The authors conclude that under this basic setup Fabric performs better, both in terms of throughput and latency.
A comparison between the native Fabric implementation and an architecture that integrates trusted chaincode execution using Intel SGX enclaves is presented by Brandenburger \emph{et~al.}~\cite{brandenburger2018blockchain}.
The use of enclaves incurs a small reduction in throughput and a small latency increase.



\vspace{-1mm}
\section{Discussion and future work}
\label{section:conclusion}

In this work, we enhanced the Fabric gossip layer.
By decreasing its latency, we decrease the number of invalidated transactions in high throughput and concurrent applications.
Our enhanced protocol also comes with a significant reduction of the overall network bandwidth consumption.
We suggest three avenues for further improvements.

Fabric does not currently allow to broadcast data blocks between peers in different organizations.
This is due to access control rules: one can evict an organization by removing it from a channel at any time, and the permission to send a block to an outside organization tightly depends on these rules.
Solving this problem requires some bookkeeping, but the development community around Fabric is considering to enable it in a future release.
This would certainly be a nice feature as Fabric scales since the good properties of epidemic algorithms shine as the number of peers increases due to the law of large numbers.

On the theoretical side, we did not consider adversarial peers trying, for instance, to hinder the dissemination by purposefully dropping blocks received from other peers.
Although gossip epidemic dissemination is obviously better than deterministic protocols in this setting, a formal analysis of the impact of adversarial peers and possible countermeasures is warranted.

The use of a stream of transactions instead of blocks, as proposed with StreamChain~\cite{istvan2018streamchain}, could reduce the latency of the ordering drastically, and put a stronger emphasis on the impact of gossip.
We keep the evaluation of enhanced gossip under this setting for future work, once StreamChain is integrated into Fabric.

We finally intend to expand the study of gossip and its impact on consistency, fairness and performance to other blockchains, both permissioned (e.g. Sawtooth~\cite{olson2018sawtooth}) and open (e.g. Ethereum~\cite{wood2014ethereum}).




\section*{Acknowledgments}

This work was partially supported by the \emph{European Union{\textquotesingle}s Horizon 2020 research and innovation programme} under grant agreement No 692178, and by The Brussels Institute for Research and Innovation (Innoviris) under project FairBCaaS.
The authors would like to thank Marco Vukoli{\'c} and Yacov Manevich from IBM Research for their valuable pointers on the inner workings of Fabric.


\bibliographystyle{plain}
\bibliography{references}



\vspace{-1mm}
\appendix

We briefly summarize how to derive the time-to-live $TTL$ required for our infect-upon-contagion push phase to disseminate a block to all peers with probability $1-p_e$ in a network of $n$ peers using a fan-out of $\fanout$. 

Let $X_r$ be the number of peers that receive at least one push digest during round $r$, and let  $\varphi(x) \triangleq n\left( 1 - \left( 1 - \frac{1}{n} \right)^{f_{\mathrm{out}} \cdot x} \right)$. Since $\varphi$ is a concave function, it follows from Jensen's inequality that $\mathbb{E}[X_{r+1}] \leq \varphi \left( \mathbb{E}[X_r] \right) = n\left( 1 - \left( 1 - \frac{1}{n} \right)^{f_{\mathrm{out}} \cdot \mathbb{E}[X_r]} \right)$. 
Applying $\varphi$ recursively we obtain $\varphi \left( \mathbb{E}[X_r] \right) \leq \phi(\phi(\E[X_{r-1}])) \leq \cdots \leq \phi^{r+1}(\E[X_0])$ where $\E[X_0] = 1$ and $\phi^{r+1}(x) = \phi(\phi^r(x))$. We define 
  $\psi : \N \rightarrow \R$ recursively by $\psi(0) = 1$ and 
    $\psi(r+1) \triangleq \phi(\psi(r)) = n\left( 1 - \left(1 - \frac{1}{n}\right)^{f_{\mathrm{out}} \cdot \psi(r)}\right)$. It follows that
  $\E[X_{r+1}] = \E[\phi(X_r)] \leq \psi(r+1)$.
$\psi(r)$ converges to a limit $\gamma$ called \emph{carrying capacity}, because it is monotonically increasing and bounded above by $n$. A solution for $\gamma$ is given by Corless et al. \cite{corless1996lambertw} with the principal branch of the Lambert-$W$ function
  $\gamma = n \frac{f_{\mathrm{out}} + W(-f_{\mathrm{out}}e^{-f_{\mathrm{out}}})}{f_{\mathrm{out}}}$  where $W(x)$ is the largest solution of $x=We^{W}$. Let $X$ be the population size at time $t$ and consider the differential equation
  $\frac{dX}{dt} = \kappa X \cdot \left( 1 - \frac{X}{\gamma} \right)$
  where $\kappa$ is the growth rate. Solving this differential equation for $X$ yields 
  $X(t) = \frac{\gamma X_0 e^{\kappa t}}{\gamma + X_0(e^{\kappa t} - 1)}$
  where $X_0 = X(0) = 1$ and $\lim\limits_{t \to \infty} X(t) = \gamma$. Since we want to compare $X(t)$ with $\psi(t)$, we choose $\kappa$ such that $e^\kappa = f_{\mathrm{out}}$, yielding
  $X(t) = \frac{\gamma f_{\mathrm{out}}^t}{\gamma + f_{\mathrm{out}}^t - 1}$. We can prove that $\psi(r) \geq X(r)$ when $\fanout \geq 2$ (we omit the proof due to lack of space).
Let $m\triangleq\sum\limits_{i = 0}^{r-1} \fanout \cdot \E[X_i]$ be the expected number of push digests transmitted during $r>1$ rounds. It follows that
$m \approx \fanout \sum\limits_{i = 0}^{r-1} \psi(i) \geq \fanout \sum\limits_{i = 0}^{r-1} X(i) \geq \fanout \int_0^{r-1} \frac{\gamma \fanout^x}{\gamma + \fanout^x - 1} dx = \gamma \fanout \log_{\fanout} \frac{\gamma+\fanout^{r-1}-1}{\gamma}$ provides a good estimate of the number of push digests that are sent on average, thus  $r \geq \log_{\fanout} \left( \gamma {\fanout}^{\frac{m}{\gamma\fanout}}-\gamma+1\right) + 1$ provides a good estimate of the number of rounds needed. 

The number of rounds required depends on the number of random digests, which in turn depends on the number of peers to infect and the probability of imperfect dissemination $p_e \leq n\left(1-\frac{1}{n}\right)^m$. Note that our analysis is conservative since it assumes that a peer can send the $\fanout$ digests to the same peer, including itself. A more precise analysis with extensions of the coupon collector's problem is possible, but does not improve the results for the networks we consider.

\end{document}